\documentclass[doublecol]{epl2}

\title{Anisotropic spin fluctuations and multiple superconducting gaps in hole-doped Ba$_{0.72}$K$_{0.28}$Fe$_2$As$_{2}$: NMR in a single crystal}
\shorttitle{Anisotropic spin fluctuations and multiple superconducting gaps in hole-doped Ba$_{0.72}$K$_{0.28}$Fe$_2$As$_{2}$} 

\author{K. Matano\inst{1} \and Z. Li\inst{1} \and  G.L. Sun\inst{2} \and D.L. Sun\inst{2} \and C.T. Lin\inst{2} \and M. Ichioka\inst{1} \and Guo-qing Zheng\inst{1,3}}
\shortauthor{K. Matano \etal}

\institute{                    
  \inst{1} Department of Physics, Okayama University, Okayama 700-8530, Japan\\
  \inst{2} Max Planck Institute, Heisenbergstrasse 1, D-70569 Stuttgart, Germany\\
  \inst{3} Institute of Physics, Chinese Academy of Sciences, Beijing 100190, China
}

\abstract{
We report the $^{75}$As-NMR study on a single crystal of the hole-doped iron-pnictide superconductor Ba$_{0.72}$K$_{0.28}$Fe$_2$As$_{2}$ ($T_{\rm c}$ = 31.5 K).   We find that the Fe antiferromagnetic spin fluctuations are anisotropic and are weaker compared to underdoped copper-oxides or cobalt-oxide superconductors.  The spin lattice relaxation rate $1/T_1$ decreases below $T_{\rm c}$ with no coherence peak and shows a step wise variation at low temperatures, 
which is indicative of   multiple superconducting gaps, as in the electron-doped Pr(La)FeAsO$_{1-x}$F$_{x}$. Furthermore, 
 no evidence was obtained for  a  microscopic coexistence of a long-range magnetic order and superconductivity.} 

\begin{document}

\maketitle


\maketitle


The discovery of superconductivity in LaFeAsO$_{1-x}$F$_x$ at the  transition temperature $T_{\rm c}$ = 26 K \cite{Kamihara} has led to a great breakthrough in the research of high temperature superconductivity. Soon after the initial work,  $T_{\rm c}$ was raised to 55 K in SmFeAsO$_{1-x}$F$_x$ \cite{Ren3}, which is the highest among materials except cuprates. 
These compounds have a ZrCuSiAs type structure (P4/nmm)  in which FeAs forms a two-dimensional network similar to the CuO$_2$ plane in the cuprates case. By replacing O with F, {\it electrons} are doped. The  superconductivity is found to be in the spin-singlet state with multiple gaps \cite{Matano}, the latter property is likely associated with the multiple electronic bands. The Fermi surface consists of  two hole-pockets centered at the $\Gamma$ point, and two electron-pockets around the M point \cite{Singh}.

After the discovery of ReFeAsO (Re: rare earth), several other Fe-pnictides have been found to superconduct.   BaFe$_2$As$_2$ has a ThCr$_2$Si$_2$-type structure, which is the very same structure for the  heavy fermion superconductor CeCu$_2$Si$_2$ \cite{Steglich}.   By replacing Ba with K, {\it holes} are doped and  $T_c$ can be as high as 38 K \cite{Rotter}.   Ba$_{1-x}$K$_x$Fe$_2$As$_2$ has attracted particular attention for the following reasons. Firstly, it  provides an  opportunity to study  possible difference between electron doping and hole doping, which have a different effect in the case of cuprates. For example, the electron correlations are usually quite strong in lightly and optimally hole-doped cuprates, while they are weak in electron-doped cuprates \cite{Zheng-LaPr}. Secondly, it has  two Fe layers in the unit cell instead of one  in ReFeAsO, which offers an opportunity to study the relationship between structure and superconductivity. Thirdly, hole doping will lead to a different evolution of the Fermi surface, which can have great impact on the superconductivity. In addition, it has been reported that antiferromagnetic order may coexist with superconductivity for 0.2$\leq x \leq$0.4 \cite{Bao}.

In this Letter, we report the first $^{75}$As-NMR (nuclear magnetic resonance) study on a single crystal of the hole-doped Ba$_{0.72}$K$_{0.28}$Fe$_2$As$_{2}$ ($T_{\rm c}$ = 31.5 K).  The measurements using single crystal allow us to address the anisotropy of electron correlations in the superconductor. We find that the Fe antiferromagnetic (AF) spin fluctuations are weaker compared to underdoped cuprates or cobaltate superconductors, and  are anisotropic in  spin space. Namely, the susceptibility at the AF wave vector is larger in the $a$-axis direction than in the $c$-axis direction. This property resembles the case of the cobaltate superconductor Na$_x$CoO$_2$$\cdot$1.3H$_2$O \cite{Matano-2} but is different from the high-$T_c$ cuprates \cite{Mali}. An NMR measurement using a powder sample was recently reported \cite{Fukazawa}, but the measured temperature was not low enough  to discuss the superconducting gap feature. We find that the spin lattice relaxation rate $1/T_1$ decreases below $T_c$ with no coherence peak and shows a broad hump at $T \sim T_c/2$,
which is an indication of multiple superconducting gaps.
The multiple-gap feature appears to be  common irrespective of the nature of doped carriers.
NMR is the most direct probe to test if the coexistence of different states takes place in the microscopic scale or not.  
We find  no evidence  for  a microscopic coexistence  of  magnetic order  and superconductivity.

The single crystals of Ba$_{1-x}$K$_{x}$Fe$_2$As$_{2}$ were grown by using  Sn-flux  \cite{Lin}. Plate-like crystals of $x$=0 and 0.28 (determined by ICP) with a surface area $\sim$8 mm$\times$ 6 mm were used for NMR measurements.  
AC susceptibility measurement using the NMR coil indicates $T_{\rm c}$=31.5 K at zero magnetic field ($H$=0). The $T_{\rm c}$ is 31 K for $H$(=7.5 T)$\|a$-axis, and 29.5 K for $H$(=7.5 T)$\|c$-axis, respectively.  
All NMR measurements were carried out by using a phase-coherent spectrometer at a fixed frequency ($f$=$\omega/2\pi$) of 55.1 MHz.  The spectrum was taken  by sweeping the magnetic field. The Knight shift ($K$) was determined with respect to $\omega/\gamma$ with $\gamma$=7.292 MHz/T.
The $1/T_1$ was measured by using a single saturation pulse.

\begin{figure}[h]
\includegraphics[width=7.5cm]{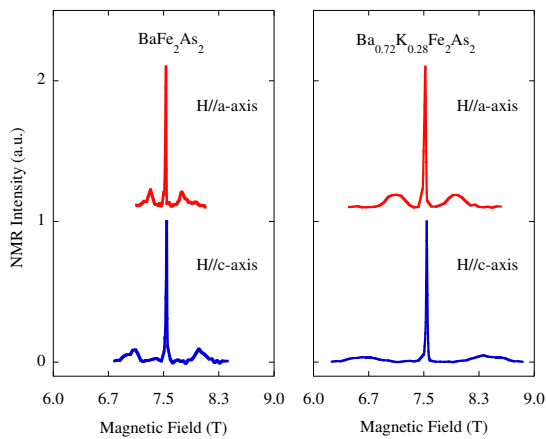}
\caption{\label{fig1} (color online)
$^{75}$As-NMR spectra at a frequency of 55.1 MHz and $T$ = 100 K for the doped and non-doped samples. The vertical axis for $H\|a$ is offset by 1.1 for clarity.}
\end{figure}

Figure 1 shows the $^{75}$As-NMR spectrum  for the parent compound and the doped sample.
The nuclear quadrupole  frequency $\nu _{\rm Q}$ for the doped sample is found to be $\sim$ 5.9 MHz at $T$ = 100 K. The $\nu _{\rm Q}$ is 3 MHz for our parent compound, which is in excellent agreement with a previous report on a similar sample \cite{Baek}.  Thus, hole doping results in a drastic increase in  $\nu _{\rm Q}$.   
 
Figure 2 shows the $T$ dependence of the Knight shift obtained from the central transition peak. 
For $H\|a$-axis, the effect of the nuclear quadruple interaction was taken into account in extracting $K^a$. 
Both $K^a$ and $K^c$ slightly decrease with decreasing $T$, but becomes a constant below $T\sim$ 70 K, which resembles closely the cuprate \cite{Zheng-LaPr} or cobaltate cases \cite{Zheng-Co}. 
\begin{figure}[h]
\includegraphics[width=8cm]{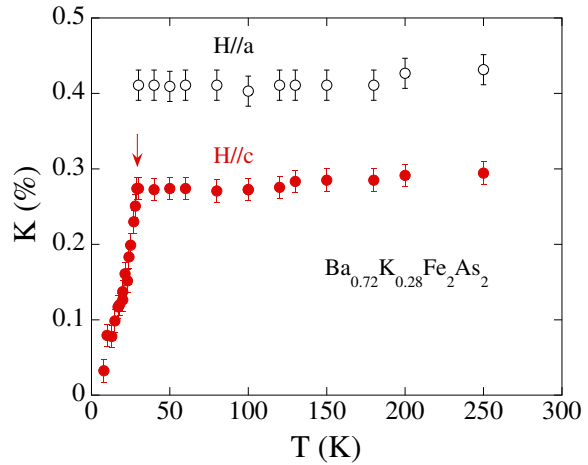}
\caption{\label{fig2} (color online) The $T$ dependence of the Knight shift.
 with $H \|$a-axis and $H \|$c-axis, respectively. The arrow indicates $T_c$.}
\end{figure}

%
Figure 3 shows the $T$ dependence of $^{75}$($1/T_1T$). The $T_1$  was measured at the central transition peak ($H$=7.5 T) and determined  from an excellent fit of the nuclear magnetization to the theoretical curve 1-$M(t)$/$M_0$ = $0.1\exp$(-$t/T_1$)+0.9$\exp$(-6$t/T_1$) (Ref.\cite{Narath}), where $M_0$ and $M(t)$ are the nuclear magnetization in the thermal equilibrium and at a time $t$ after saturating pulse, respectively. For comparison, the data for  BaFe$_2$As$_2$ ($T_N$=80 K) were also plotted in the figure, which are in agreement with previous reports \cite{Baek,Kitagawa}. 
\begin{figure}[h]
\includegraphics[width=8cm]{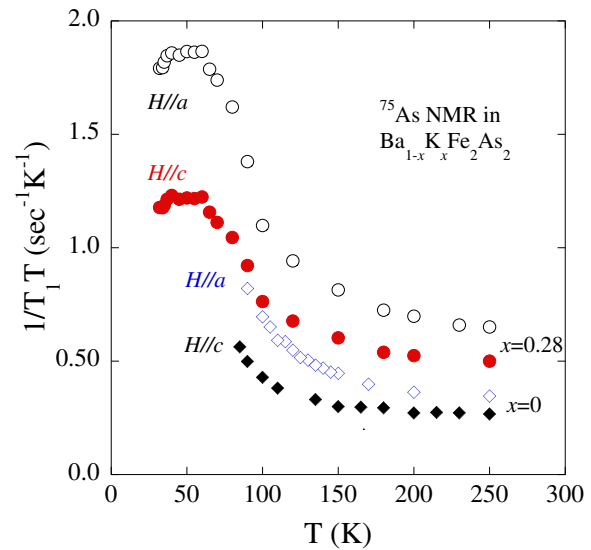}
\caption{\label{fig3} (color online)
The quantity $^{75}$($1/T_1T$) in the normal state of Ba$_{0.72}$K$_{0.28}$Fe$_2$As$_{2}$ (circles) and in the paramagnetic state of BaFe$_2$As$_{2}$ (diamonds). }
\end{figure}

We first discuss the electron correlations and their anisotropy above $T_c$. 
The Knight shift is composed of two parts, the part $K_s$ due to spin susceptibility $\chi_s$ and $K_{orb}$ due to  orbital susceptibility. Namely, $K= K_s+K_{orb}$ and  $K_s=A_{hf}(0)\chi_s$,
where $A_{hf}$(0) is hyperfine coupling constant at wave vector $q$=0.
As can be seen in Fig. 2, below $T_c$, $K^{c}$ decreases and approaches to zero, which is consistent with the  spin-singlet pairing symmetry and suggests $K_{orb}^c \sim$0. The spectrum for $H\| a$-axis becomes much broader below $T_c$ which makes the determination of $K^a$ more difficult and will be discussed in a separated publication in the near future. Meanwhile, it is noted that
 $K_{orb}^a$ is also negligible in the electron-doped counterpart \cite{Matano,Grafe}. Thus the anisotropy in $K$ is likely due predominantly to the spin part $K_s$, and can be accounted for by the anisotropy of the hyperfine coupling constant, which was found to be  $\frac{A_{hf}^a(0)}{A_{hf}^c(0)}$=1.4 in  BaFe$_2$As$_2$ \cite{Kitagawa}.

From Fig. 3 one notices that,  in the normal state above $T_c$,
$1/T_1T$ increases with decreasing $T$, although not as steeper as in the paramagnetic state of BaFe$_2$As$_2$. This is an indication of antiferromagnetic (AF) electron correlation, since the spin susceptibility at $q$=0 does not increase upon cooling. However, below  $T\sim$ 70 K, $1/T_1T$ becomes $T$-independent.  This indicates that a modified Korringa relation is satisfied below this temperature and that the system is in a renormalized Fermi liquid state. This is quite different from 
the cases of underdoped cuprates \cite{Ohsugi} or cobalt-oxide superconductor \cite{Fujimoto} where  upon cooling antiferromagnetic correlations develop all the way down to $T_c$. In this context, the spin correlations are weaker in the present compound than the cuprates or cobaltates. This property seems to be common for Fe-pnictides irrespective of hole- or electron-doping \cite{Matano}.

In a general form, $1/T_1T$  is written as
\begin{eqnarray} 
\frac{1}{T_1T}= \frac{\pi k_B \gamma^2_n }{(\gamma_e \hbar )^2} \sum_q A_{hf}(q)^2 \frac{\chi ''_{\perp}(q,\omega_n)}{\omega_n},
\end{eqnarray}
 where 
$\chi ''_{\perp}(q,\omega_n)$ is the imaginary part of the dynamical susceptibility  perpendicular to the applied field.
The larger magnitude of $1/T_1T$ along the $a$-axis direction  than that along the $c$-axis direction indicates that there are stronger fluctuations along the $c$-axis direction seen by  the As-site, as discussed in detail below. 


%
\begin{figure}[h]
\includegraphics[width=8cm]{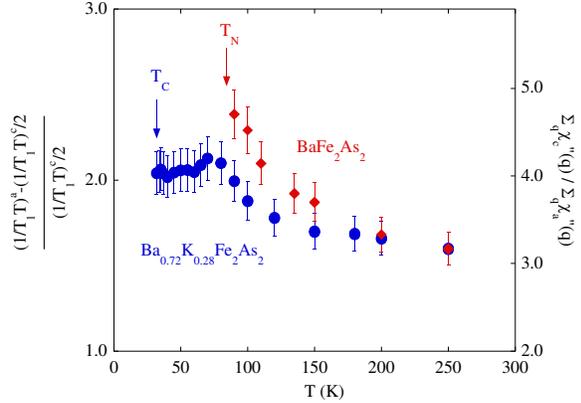}
\caption{\label{fig:kai} (Color online) $T$-dependence of the anisotropy of the spin fluctuations seen at the As site in terms of $\frac{\sum_{q} A_{hf}^c(q)^2 \chi ''_c \left(q\right) }{\sum_{q} A_{hf}^a(q)^2 \chi ''_a \left(q \right)}$ (left axis) and $\frac{\sum_{q}  \chi ''_c \left(q\right) }{\sum_{q}  \chi ''_a \left(q \right)}$ (right axis).}
\end{figure}
 If we neglect the planar anisotropy, each component of  $\left( 1/T_1T \right)$  is related to  $\sum_{q} \chi '' (q)$ through the following expressions,
\begin{eqnarray} 
\left(\frac{1}{T_1T} \right)^c= 2\frac{\pi k_B \gamma^2_n }{(\gamma_e \hbar )^2 \omega_n} \sum_{q} A_{hf}^a(q)^2 \chi ''_a \left(q \right) \\ 
\left(\frac{1}{T_1T} \right)^a= \frac{\pi k_B \gamma^2_n }{(\gamma_e \hbar )^2 \omega_n} \sum_{q} \left[ A_{hf}^a(q)^2 \chi ''_a \left(q \right)+A_{hf}^c(q)^2 \chi ''_c \left(q\right) \right]
\end{eqnarray}
Then, we have the ratio
\begin{eqnarray} 
 \frac{\sum_{q} A_{hf}^c(q)^2 \chi ''_c \left(q\right) }{\sum_{q} A_{hf}^a(q)^2 \chi ''_a \left(q \right)}=  \frac{\left(\frac{1}{T_1T} \right)^a-\frac{1}{2}\left(\frac{1}{T_1T} \right)^c}{\frac{1}{2}\left(\frac{1}{T_1T} \right)^c}
\end{eqnarray}
This quantity is plotted in Fig. 4. The data for the parent compound is also shown for comparison.
The ratio in Ba$_{0.72}$K$_{0.28}$Fe$_2$As$_{2}$ is around 1.5 at high $T$ but increases to 2 below 100 K. Also, the $T$ dependence of the ratio is reduced uppon doping hole. The $A_{hf}(q)$  seems to be $q$-independent. 
  By using $A_{hf}^c$=1.88 T/$\mu_B$ and $A_{hf}^a$=2.64 T/$\mu_B$ obtained for the parent compound \cite{Kitagawa}, the anisotropy of the spin fluctuation  $\frac{\sum_{q}  \chi ''_c \left(q\right) }{\sum_{q}  \chi ''_a \left(q \right)}$ seen at the As site is  shown in Fig. 4 with scale on the right axis.

Now, neutron experiment found  that, in the undoped BaFe$_2$As$_2$ compound, the ordered Fe magnetic moment is along the $a$-direction and forms a stripe \cite{neutron}. Since the As atom sits in the position above (below) the middle of four Fe-atoms, the above results imply that, in the Fe site, a stronger fluctuating field exists along the $a$-axis direction, as illustrated in  Fig. 5.

\begin{figure}[h]
\includegraphics[width=6cm]{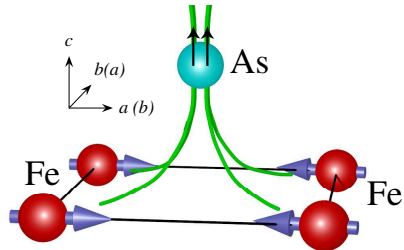}
\caption{\label{fig:kai} (Color online) Sketch of the Fe-As block. The arrows   illustrate  the larger component of the  fluctuating field of Fe  and that seen by the As site. The indicated crystal axes are defined for the orthorhombic phase of the parent compound. In the paramagnetic (tetragonal) phase, the $a$-axis is defined as rotated by 45 $^o$.}
\end{figure}

It is remarkable  that the Fe antiferromagnetic fluctuations  are anisotropic in  spin space. Namely, 
 $\chi ''_{\pm}(Q)$ is  much larger than $\chi ''_{zz}(Q)$, where $z$ is along the $c$-axis direction. This is in contrast to the high-$T_c$ cuprates where the difference between the spin fluctuations along the $c$-axis and the $ab$-plane is small \cite{Mali,Shirane}, but similar situation was encountered in cobaltate superconductor \cite{Matano-2}. The relationship between the energy- and $q$-dependences of the spin fluctuations (SF) and possible SF-induced superconductivity  has been studied both theoretically \cite{Moriya} and experimentally \cite{ZhengTl}.  To our knowledge, however, the relationship between the anisotropy of SF and superconductivity  has  been less explored so far. In the three-dimensional Hubburd Model, it was suggested that an anisotropic SF is un-favored for superconductivity compared to isotropic SF \cite{Scalapino}. We hope that our results  will stimulate  more theoretical work in this regard. 

\begin{figure}[h]
\includegraphics[width=8cm]{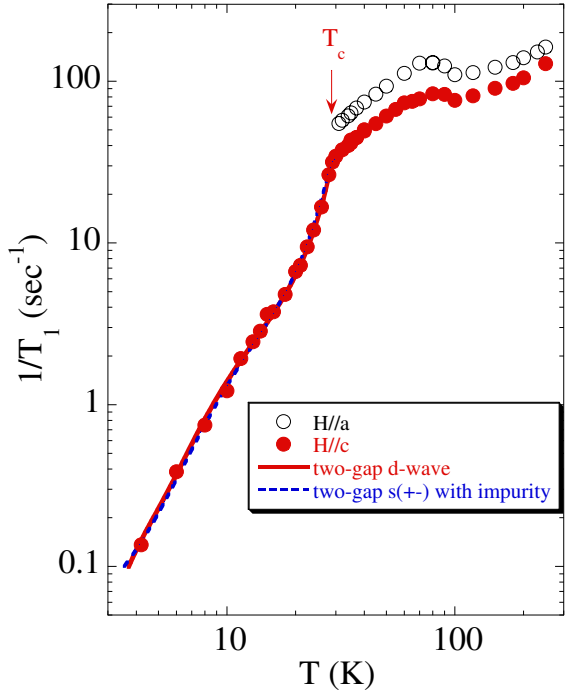}
\caption{\label{fig4} (color online) $T$-dependence of 1/$T_1$. 
%
The curves below $T_c$ (indicated by the arrow) are fits to  two-gap models  (see text).  }
\end{figure}

We next move on to discuss  the superconducting state. As seen in Fig. 6, 
$1/T_1$ decreases with no coherence (Hebel-Slichter) peak just below $T_{\rm c}$, which is in contrast to conventional BCS superconductors with an isotropic gap.  Obviously, a $d$-wave gap will suppress the coherence peak just below $T_c$. The $s^{\pm}$-symmetry with two $s$-wave gaps that change sign on different Fermi surfaces, which was recently proposed for the iron-pnictides \cite{Mazin,Kuroki}, may also reduce the coherence peak provided that the scattering between the hole- and electron-like bands is strong enough. 
Upon further cooling, however, the $T$ dependence  is not a simple power law, such as $1/T_1$ $\propto$ $T^3$ (Ref.\cite{Zheng,Asayama}) or $1/T_1$ $\propto$$T^5$ (Ref.\cite{Katayama})  as seen in heavy fermion compounds  or high-$T_{\rm c}$ cuprates, nor exponential as seen in conventional BCS superconductors.

The most striking feature is that $1/T_1$ shows a "knee"-shape around $T\sim$0.5 $T_c$. Namely, the sharp drop of $1/T_1$ just below $T_{\rm c}$ is gradually replaced by a slower change below $T\sim$ 15 K, then followed by another steeper drop below.
 This "convex" shape is clearly different from  the case of  superconductors with a single  gap which show a "concave" shape of $T$-variation. 
 It should be emphasized that this unusual $T$-variation is not due to sample inhomogeneity, which would result in a two-component $T_1$. We find that $T_1$ is of single component throughout the whole $T$ range. Such  peculiar feature  was first found in the electron-doped PrFeAsO$_{0.89}$F$_{0.11}$  \cite{Matano}
 and confirmed in  LaFeAsO$_{0.92}$F$_{0.08}$ \cite{Kawasaki}, and was interpreted as due to multiple gaps,
This  finding  was subsequently echoed by many other experiments \cite{Weyenech,Ding}.

In the present case, a two-gap model can reproduce the step-wise $T$ variation of $1/T_1$. 
The underlying physics is that the system 
is dominantly governed by a larger gap for $T$ near $T_c$ while at sufficiently low $T$  it starts to "notice" the existence of a smaller gap, resulting in another drop $1/T_1$ below 10 K. 
%
In the $d$-wave case with two gaps, where the density of states (DOS) is $N_{s,i}(E)$ = $N_{0,i}$$\frac{E}{\sqrt{E^2-\Delta_i^2}}$, $1/T_{1s}$ in the superconducting state is written as
$\frac{T_{1N}}{T_{1s}}= \sum_{i=1,2}{  \frac{2}{k_BT} \int \int N_{s,i}(E)N_{s,i}(E') f(E)\left[ 1-f(E') \right] \delta(E-E')dEdE' }$
, where $f(E)$ is the Fermi distribution function. 
We find that the parameters 2$\Delta_1(0) = 9.0 k_{\rm B}T_{\rm c}$, 2$\Delta_2(0) = 1.62 k_{\rm B}T_{\rm c}$  and  $\kappa = 0.69$ can fit the data very well as shown by the solid curve in Fig. 6, where 
\begin{eqnarray}
\kappa = \frac{N_{0,1}}{N_{0,1}+N_{0,2}} 
\end{eqnarray}
is the relative  DOS of the band(s) with  larger gap to the total DOS.  Comparison with electron-doped PrFeAsO$_{0.89}$F$_{0.11}$, where $2\Delta_1(0)$ = 7.0 $k_{\rm B}T_{\rm c}$, $2\Delta_2(0)$ = 2.2$k_{\rm B}T_{\rm c}$, and $\kappa$ = 0.4  within the same model \cite{Matano}, shows that the $\Delta_1$ is substantially larger in the hole-doped case. 
The same parameters can also fit the Knight shift data as shown  in  Fig. 7.
\begin{figure}[h]
\includegraphics[width=7.5cm]{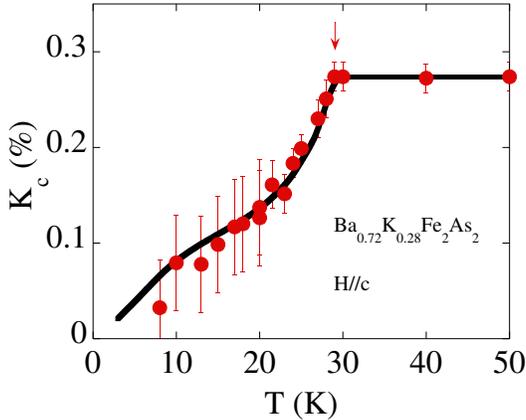}
\caption{\label{fig2} (color online) A blow-up of the low-$T$ Knight shift data for $H \|$c-axis. The arrow indicates $T_c$. The curve below $T_c$ is a fit to a two-gap model with the same parameters used for fitting $T_1$ (solid curve in Fig. 6).}
\end{figure}

For the case of $s^{\pm}$-gap, recent calculations have shown that scattering between the  different bands may reduce the coherence peak just below $T_{\rm c}$ \cite{Chubukov,Bang}. Following Ref.\cite{Bang}, we calculated $1/T_1$ for the $s^{\pm}$-gap model, by introducing the impurity scattering parameter $\eta$ in the energy spectrum, $E=\omega+i\eta$. The parameters 2$\Delta^{+}_1(0) = 7.2 k_{\rm B}T_{\rm c}$, 2$\Delta^{-}_2(0) = 1.67 k_{\rm B}T_{\rm c}$,    $\kappa = 0.6$ and $\eta$=0.22$k_{\rm B}$$T_{\rm c}$  can well fit  the data, as shown in  Fig. 6 \cite{Note}.  


It is worthwhile pointing out that $\Delta_1$ is in the strong coupling regime, while the $\Delta_2$ is much smaller than the BCS value. Angle resolved photoemission spectroscopy found that the gap on the  inner hole-pocket ($\alpha$-band) and the  electron-pockets ($\gamma$, $\delta$-bands) has a larger value of $2\Delta=7.2\sim 7.7k_BT_c$, while the gap on the  outer hole-pocket ($\beta$-band) is smaller \cite{Ding}. It is tempted to assign the larger gap $\Delta_1$ found in the present work to  the gap on $\alpha$-band and the electron pockets, while the smaller gap $\Delta_2$ to the $\beta$-band. 
The large value of $\Delta_1$ could be attributable to the excellent Fermi-surface nesting which can enhance pairing in the $s^{\pm}$-wave scenario, while  the small value of $\Delta_2$  would be owing to   the lack of nesting counterpart of the $\beta$-band. In the electron-doping case, on the other hand, the $\beta$-pocket shrinks so that it becomes to nest with the electron pockets but the nesting is not as good as the hole-doped case \cite{Ding2}. This may explain the smaller $\Delta_1$ in  the electron-doped PrFeAsO$_{0.89}$F$_{0.11}$.

Finally we   address the issue of possible coexistence of antiferromagnetic order with superconductivity.
%
Neutron scattering experiment  suggested that, in the doping range of 0.2$\leq x \leq$0.4,  a long-range antiferromagnetic order   coexists with superconductivity \cite{Bao}.  However, our result shows that such AF order, if exists, is  macroscopically phase-separated from the main phase that undergoes the superconducting transition. This is because no internal magnetic field arising from the AF order is detected in the NMR spectrum of the superconducting phase, nor shows any  anomaly attributable to an AF order in the $T$ dependence of $1/T_1$.  Below $T$=70 K,  as mentioned already,  the system satisfies the Korringa relation  expected for a weakly correlated electron system. 
We do observe an anomaly at $T_{0}\sim$100 K in $1/T_1$, but this $T_{0}$ is higher than $T_N$= 80 K of the parent compound. A similar anomaly was found by Baek {\it et al} \cite{Baek} at 140 K in BaFe$_2$As$_2$. The origin is unclear at the moment.

In summary, we have presented the  NMR results on  the hole-doped iron-pnictide single crystal Ba$_{0.72}$K$_{0.28}$Fe$_{2}$As$_{2}$ ($T_{\rm c}$ = 31.5 K). We find a weak antiferromagnetic spin fluctuation that is anisotropic in  spin space.  The  $1/T_1$ decreases below $T_{\rm c}$ with no coherence peak and does not follow a simple power-law nor exponential function. As in   the electron-doped (Pr,La)FeAsO$_{1-x}$F$_x$, the result indicates multiple superconducting gaps. However,  $\Delta_1$ is substantially greater than the corresponding value in the electron-doped counterparts.
We find   no evidence  for a microscopic  coexistence of magnetic order and superconductivity. 

We thank  I. Harada, H. Kontani, K. Kuroki and Kazuo Ueda for useful discussion, H. Fukazawa for a helpful comment, and S. Kawasaki for help in some of the measurements.
This work was supported in part by research grants from MEXT and  JSPS (No. 20244058 and No. 17072005).



\begin{thebibliography}{}
\bibitem{Note}
A three-band model with 2$\Delta^+_1(0)=7.2 k_{\rm B}T_{\rm c}$, 2$\Delta^+_2(0)=1.67 k_{\rm B}T_{\rm c}$, 2$\Delta^{-}_3(0)=7.2 k_{\rm B}T_{\rm c}$ and $N_{0,1}$:$N_{0,2}$:$N_{0,3}$=0.3:0.4:0.3 can also fit the data well.
\end{thebibliography}

\end{document}